\documentclass[twocolumn,showpacs,preprintnumbers,amsmath,amssymb]{revtex4}
\usepackage{graphics}
\begin{document}
\title{The Foam Analogy in Charged Colloidal Crystals}
\author{William Kung}
\affiliation{Department of Physics and Astronomy, University of
Pennsylvania, Philadelphia, PA 19104-6396}
\author{P. Ziherl}\altaffiliation{Permanent address: J. Stefan
Institute, Jamova 39, SI-1000 Ljub\-lja\-na, Slovenia}
\author{Randall D. Kamien}\email{kamien@physics.upenn.edu}
\affiliation{Department of Physics and Astronomy, University of
Pennsylvania, Philadelphia, PA 19104-6396}
\date{\today}
\begin{abstract}
We model charged colloidal suspensions using an analogy with
foams. We study the solid--solid phase transitions of
these systems as a function of particle volume-fraction and ionic
strength. The screened Coulomb interaction is replaced by an interaction
between walls of the Voronoi cells around each particle.  We fit the surface
charge to reproduce
the phase diagram for the charged suspension
studied by Sirota {\sl et al.} \protect{[\prl {\bf 62}, 1524 (1989)]}.
With this fit parameter we are able to calculate the elastic moduli of the system
and find good agreement with the available data.
\end{abstract}
\pacs{61.50.Ah, 62.20.Dc}
\maketitle

Charged colloids have been the subject of intense study both
experimentally and theoretically. In the laboratory, optical techniques
can readily probe these systems and they are easily manipulated via
chemical means~\cite{Hachisu73,Yoshiyama86,Russel89,Mono89}. Their
rich chemistry leads to many industrial applications ranging from uses
in coating materials, in ceramic precursors, and in designing and
manufacturing biological macromolecules. These systems interact
through a short-range screened Coulomb potential and thus epitomize one
of the few classes of interactions in soft systems.
In this Letter, we replace the exact screened-Coulomb interaction
between colloidal spheres with an effective interaction proportional
to the surface area of the Wigner-Seitz or Voronoi cells which contain
each sphere. By balancing the entropic interaction which favors the
close-packed face-centered cubic (FCC) lattice with the surface interaction
which, as we will argue, favors the body-centered cubic (BCC) lattice, we
will find the phase coexistence line as a function of volume fraction and
salt concentration.  We compare our results to experiment to find the surface charge on the
spheres.  Using this fitted value, we calculated the shear moduli within our framework and
found close agreement with experiment. Our model leads to quantitatively different results than
those found in molecular dynamics (MD) simulations~\cite{Robbins88} of particles
interacting via Yukawa interactions.  

Several experimental studies~\cite{Hachisu73,Yoshiyama86,Russel89}
provide a wealth of data on the stability of colloidal phases,
in particular, the fluid phase, the glass phase and two solid phases
with either FCC or BCC lattices. These systems are aqueous suspensions of
uniformly charged polystyrene spheres with a variable salt
concentration which is a control parameter for the degree of
screening of the underlying Coulomb interaction. Because the lattice
spacing is large in the ordered phases, they can be probed 
by scattering of visible light~\cite{Mono89}. Through
scattering it is possible to determine those volume fractions at which
the order-disorder transition occurs at low salt concentrations and at
which the FCC-BCC transition occurs.  Though the MD simulations~\cite{Robbins88}
qualitatively corroborate these
experimental findings, one
is hard-pressed for similar quantitative agreement between existing
experimental data and simulation results. The lack of exact analytic
methods with which to compare experiment or simulation further
complicates this problem. Thus theoretical models which distill the
crucial aspects of the interactions and lead to testable predictions
are necessary.

Recently, a foam analogy was used to account for the many non-closed
packed structures observed in colloidal and lyotropic
systems~\cite{Ziherlshort,Ziherllong}. In this analogy, the surface interaction
replaced the brush-like interaction between the spheres and accounted
for the entropic degrees of freedom associated with configurations of
the soft coronas of alkyl chains attached to hard cores. Since the
coronas prefer a packing arrangement that disfavors interdigitation and
thus maximize the average separation of the cores, lattices with
small interfacial area are favored in such systems. For example, the BCC
lattice of orthic tetrakaidecahedra has a smaller area than the FCC
lattice of rhombic dodecahedra, and so the surface interaction
favors the BCC lattice.

Here we apply the same framework to understanding the experimental phase
diagram~\cite{Sirota89} and MD simulation results~\cite{Robbins88} for
charged colloidal systems. In particular, our model couples the bulk
free energy of the charged hard cores to the free energy of their
screened Coulomb interaction via the effective surface area. The model
contains a single free parameter: the surface potential associated with
each colloidal particle. Using experiment as our guide, we fit
the surface potential to reproduce the phase diagram. In order to test
our theory, we use the same surface potential to calculate the shear and
bulk moduli and compare these with experiment and simulation.  We
have found good agreement with the available data and, in doing so,
demonstrate the plausibility of understanding the various experimental
results of charged systems in terms of a geometrical principle already
established for another class of colloids.

An apparent dramatic feature of our results is the negative values
of the bulk modulus for small $\lambda=\kappa a$, where $\kappa$ is
the inverse Debye screening length and $a$ is the average interparticle spacing. This is an artifact of
our model which can be simply understood: for $\lambda=1$ the
Debye screening length is exactly equal to the interparticle
spacing. As a result, we would expect next-to-nearest-neighbor
interactions to play an important role in the stability and
energetics of the charged colloidal crystal and therefore would
not trust our model at small $\lambda$.  However, for larger
values of $\lambda$ the bulk modulus become positive -- for the
data we compare with, $\lambda\approx 4.64$, well within the
regime where next-to-nearest-neighbor interactions are small. Our
model also yields qualitatively different behavior for the
relative magnitudes of the elastic constants in comparison to MD
simulations~\cite{Robbins88}. We attribute this difference to the
finite size of the colloids -- in the MD simulation they were
treated as point particles and thus did not suffer entropic losses
from excluded volume. We also note that the shear modulus becomes
negative at large values of $\lambda$. This signals the
instability of the BCC lattice and its transition to FCC: for
large $\lambda$ the screened-Coulomb potential becomes unimportant
and the system should behave as if the interactions were purely
hard-core. This is in line with the recent experimental
determination~\cite{Grier98} that the FCC lattice is the
equilibrium phase for samples with $\lambda>10$. Moreover, in that
study it was argued that since Yukawa-like pair interactions could
not account for both the elasticity of the crystal and the
colloidal dynamics, three-body and higher interactions must play
an important role. Since our approach is intrinsically many-body,
our results for the elastic constants as a function of $\lambda$
are consistent with that conclusion -- we find behavior that
cannot be explained by MD simulation of particles interacting via
Yukawa potentials.  There have been alternative explanations of
this discrepancy based on hydrodynamic screening \cite{Brenner}.
We hope that our framework will allow for improved understanding
of these issues.

In a colloidal suspension at fixed density, the volume of the
container is the sum of the volume of the hard spheres and the excess
volume of the salt solution. Since this latter volume can be viewed
as enveloping the individual spheres, we can imagine breaking the
volume up into a lattice of Voronoi cells, each of which contains a
colloidal particle. The excess volume can then be written as the
product of the area of these dividing surfaces and their average thickness
so that
\begin{equation}
Ad={\rm constant},
\label{Ad}
\end{equation}
where $A$ is total area of these bilayers and $d$ is their average
thickness. Maximizing $d$ to reduce the repulsive interaction
amounts to minimizing $A$ whence comes the
interaction that favors BCC lattices (and the more exotic A15
lattices~\cite{Ziherlshort,Balagurusamy}). Since the crystal lattice is
composed of identical colloidal particles in cases of interest, the
minimum area problem is equivalent to Kelvin's problem~\cite{Thomson1887} of
partitioning a given space into cells of equal volumes having the
smallest interfacial area.

To compute the bulk free energy of the system, we follow the method
developed in a previous study~\cite{Ziherlshort,Ziherllong} by
adopting the cellular free-volume theory. In this approximation, each
particle is confined to a cage formed by its neighbors. The free
volume available to each particle's center of mass is the volume of
the Wigner-Seitz cell after a layer of thickness $\sigma/2$ (where
$\sigma$ is the hard-core diameter of particles) is peeled off of its
faces. Despite it mean-field nature, the free volume theory yields
excellent quantitative agreement with available numerical simulations
in the high-density limit and is roughly $1\%$ off above the hard-sphere
fluid-solid transition.   The use of free volume theory at the lower densities probed in
the charged colloidal systems could be suspect.  However as long as the shear elastic constants
are non-zero,
we expect that an ``Einstein-crystal'' description of the phonon modes should be
adequate.  Indeed, we will find that appropriate moduli are non-zero in the density and salt concentration regime we
are studying \cite{Robbinspers} and therefore there are no ``soft modes'' which might strongly
contribute to collective effects.
The resulting bulk free energy
for the FCC or BCC lattice is:
\begin{equation}
F^{\scriptscriptstyle X}_{\rm bulk}=-k_{\scriptscriptstyle
B}T\ln\bigglb(\alpha^{\scriptscriptstyle X}\left(\beta^{\scriptscriptstyle
X}n^{-1/3}-1\right)^3\biggrb),
\label{Fbulk}
\end{equation}
where $n=\rho\sigma^3$ is the reduced density and $\sigma$ is the
hard-core diameter of each colloidal particle. The
coefficients $\alpha^{\scriptscriptstyle FCC}=2^{5/2}$ and
$\alpha^{\scriptscriptstyle BCC}=6.716$ depend on the shape of the
cells, whereas $\beta^{\scriptscriptstyle FCC}=2^{1/6}$ and
$\beta^{\scriptscriptstyle BCC}=2^{-2/3}3^{1/2}$ are determined by
their size.

To obtain the electrostatic energy between the charged spheres, we
model their interaction as that between two charged flat plates using
the Debye-H\"uckel approximation~\cite{Gast83}\ for a salt solution with counterions of
charge $Ze$. We replace each sphere
with its Voronoi cell and then, in a Derjaguin-like approximation, replace the
interaction with an interaction across parallel plates.
The energy between two
plates can be found via an electromagnetic potential satisfying the
linearized Poisson-Boltzmann equation. For the experimental system in
question we will show that this approximation is valid. Using linear
superposition and the solution for a single charged plate, we
have~\cite{Russel89}:
\begin{equation}
F_{\rm c}=64A_{\rm M}k_{\scriptscriptstyle B}Tn_{\rm b}\kappa^{-1}{\rm
tanh}^2\left({1\over4}\Psi_{\rm s}\right)\exp\left(-\kappa d\right).
\label{Finterfacial}
\end{equation}
Taken together, Eqs. (\ref{Fbulk}) and (\ref{Finterfacial}) encode the
complete picture of the structural FCC-BCC transition in terms of the
bulk counterion number density $n_{\rm b}$, the dimensionless surface
potential of colloids $\Psi_{\rm s}$, the colloid density $n$, and the
Debye screening length $\kappa^{-1}=\sqrt{\epsilon k_{\scriptscriptstyle B}
T/2e^{2}Z^{2}n_{b}}$,
itself a function of these control variables (where $\epsilon$ is the dielectric constant).
In order to relate the
interaction energy $F_{\rm c}$ to the reduced density $n$, we relate
the spacing of the charged plates to the density via the constraint
Eq.~(\ref{Ad}). The interparticle separation $d$ depends on the particle
density:
\begin{equation}
A_{\rm \scriptscriptstyle M}d=2\left(\frac{1}{n}-\frac{\pi}{6}\right)
\sigma^{3},
\label{leftovervolume}
\end{equation}
where $A_{\rm \scriptscriptstyle M}=\gamma^{\scriptscriptstyle X}
\sigma^2n^{-2/3}$, and $\gamma^{FCC}=5.345$ and $\gamma^{BCC}=5.308$ are dimensionless
quantities characterizing the magnitude of area. Substitution of
Eq.~(\ref{leftovervolume}) into Eq.~(\ref{Finterfacial}) completes the
expression for the interfacial free energy. Since the
density and salt concentration are measurable parameters, the only free
parameter is the surface potential $\Psi_{\rm s}$. It is therefore
customary to treat $\Psi_{\rm s}$ as a parameter chosen for the best
fit to the data~\cite{Gast83}. From the surface potential we will be
able to determine the net surface charge $q$ on the colloidal spheres.

\begin{figure}[htb]
\vspace{2mm}
\hspace{23mm}
\includegraphics{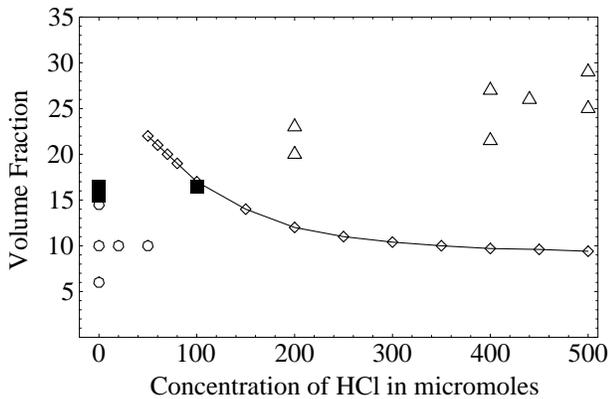}
\vspace{2mm}
\caption{Theoretical FCC-BCC coexistence curve as a function of
volume fraction $\phi$ and electrolyte concentration (HCl). The
diamonds are the theoretical predictions, while the other points come
from the data of Ref.~\protect\cite{Sirota89}. Solid squares are
coexistence points, open triangles are FCC, and open circles are BCC.}
\label{phasediagram}
\end{figure}

Focusing on the structural phase transition from BCC to FCC, we rely on
the X-ray scattering study of charged polystyrene spheres by Sirota
{\sl et al.}~\cite{Sirota89}. That experiment studied the phase diagram
of polystyrene spheres in a 0.9-methanol--0.1-water suspension. The thermodynamic
behavior of this system was recorded as a function of volume fraction
$\phi$ and salt concentration $c_{\rm HCl}$. In order to determine
the surface potential, we choose an FCC-BCC coexistence point in the
experimental data, and then equate the corresponding FCC and BCC free
energies to solve for $\Psi_{\rm s}$. Through Gauss's law, we
calculate the charge per colloid $q=2A_{M}\sqrt{2\epsilon
k_{\scriptscriptstyle B}Tn_{b}}\;{\rm sinh}\left(\frac{1}{2}\Psi_{\rm
s}\right)$.  Finally, using this value of the charge, we find the density
$n$ at which $F^{FCC}/k_{\scriptscriptstyle B}T$ and $F^{BCC}/k_{\scriptscriptstyle
B}T$ are equal for each salt concentration $n_{\rm b}$. Due to the
limited data, there is only one experimental point at non-zero salt concentration
that is in the coexistence region. We thus obtain the coexistence
curve in ($\phi$-$n_{\rm b}$) phase diagram
(Fig.~\ref{phasediagram}). The relevant experimental data from
Ref.~\cite{Sirota89} are included for comparison.

Within our framework we find a surface potential of $\Psi_{\rm s}=0.2$,
and so we were justified in linearizing the Poisson-Boltzmann equation.
However, this potential corresponds to a total charge of about $48e$
per colloidal particle. The value of the charge is smaller than the
quoted experimental value of about $135e$ per sphere~\cite{Sirota89},
though it is comparable in magnitude. The experimental value was actually
determined indirectly by measuring the shear modulus for the
crystalline sample at $12\%$ volume fraction~\cite{Joanny}. Because of
the uncertainty in the charge, we will make a direct comparison with
experiment by calculating the shear modulus of the BCC lattice using
our cellular framework.

In general, the elastic energy of a cubic crystal has three elastic
constants, $K_{11},K_{12},$ and $K_{44}$~\cite{Lubenskytext}:
\begin{eqnarray}
F_{\rm cubic}=\frac12\int {\rm d}^3x\Big[K_{11}\left(u^2_{xx}
+u^2_{yy}+u^2_{zz}\right)\nonumber\\
+K_{12}\left(u_{xx}u_{yy}+u_{xx}u_{zz}+u_{yy}u_{zz}\right)\nonumber\\
+2K_{44}\left(u^2_{xy}+u^2_{xz}+u^2_{yz}\right)\Big].
\label{elasticfree}
\end{eqnarray}
For a polycrystalline sample, the bulk modulus
$K=\left(K_{11}+2K_{12} \right)/3$ is an isotropic quantity. On
the other hand, the shear modulus depends on the direction of the
applied shear and ranges from the elongational shear
$\mu=K_{11}-K_{12}/2$ to the simple shear $K_{44}$. We determine
the three elastic constants by calculating the free volume and the
change in surface area for a deformed BCC lattice. The free volume
was found via a simple numerical scheme while the change in
surface area was found with the aid of Surface
Evolver~\cite{Brakke}. Equating the continuum energy change from
Eq.~(\ref{elasticfree}) with the energy change within our
framework, we find that $K_{11}\approx17.1$~N/m$^2$,
$K_{12}\approx -1.2$~N/m$^3$, and $K_{44}\approx10.3$~N/m$^2$ for
a $12\%$ volume-fraction ($n=0.23$) BCC sample. Thus the shear
modulus ranges from 10.3 to 17.7~N/m$^2$. Again this is the same
order of magnitude as the measured, isotropic value of
17~N/m$^2$~\cite{Sirota89}, which is a complicated combination of
the two shear moduli.  We have calculated these moduli using an HCl concentration
of 50 $\mu$mol, as this is in the middle of the reported BCC regime~\cite{Sirota89}.

In order to compare with MD simulations~\cite{Robbins88}, we tabulate
(Table~\ref{elastictable}) values of the bulk modulus and the three
elastic constants as a function of $\lambda=\kappa a$ (in the experimental study, $\lambda\approx
4.64$). However, unlike point particles with Yukawa interactions, our
model does not scale simply with this parameter. To vary $\lambda$, we
keep the colloidal density fixed as well as the total charge on each
colloid. Varying the salt concentration changes both $\kappa$ and the
surface potential $\Psi_{\rm s}$. With these new values we can
recalculate the elastic constants. For completeness, we tabulate the
values of the bulk modulus $K$ and elastic constants of the BCC
structure at $\phi=0.12$ for different values of $\lambda=\kappa a$
(where $a$ is the mean separation of spheres; at $n=0.23$,
$a\approx140$~nm):
\begin{table}
\begin{tabular}{lcccccccl}

$\lambda$& $K$&\quad&$K_{11}$&\quad&$K_{12}$&\quad&$K_{44}$\\ \hline
1 & $-26.1$ / $-7.59$ && 17.5 / 10.7 && $-47.9$ / $-16.7$ && 54.4 / 28.4 \\
2 & $-7.54$ / 13.8 && 19.0 / 18.4 && $-21.1$ / 11.6 && 29.1 / 15.6 \\
3 & 0.12 / 22.3 && 19.7 / 21.4 && $-9.67$ / 22.7 && 18.3 / 10.6 \\
4 & 3.66 / 26.2 && 18.4 / 22.4 && $-3.71$ / 28.0 && 12.7 / 7.95 \\
5 & 5.37 / 28.0 && 16.6 / 22.5 && $-0.22$ / 30.7 && 9.40 / 6.40 \\
6 & 6.16 / 28.6 && 14.5 / 22.0 && 2.00 / 32.0 && 7.30 / 5.43 \\
7 & 6.46 / 28.7 && 12.5 / 21.1 && 3.40 / 32.5 && 5.90 / 4.70 \\
8 & 6.49 / 28.5 && 10.6 / 20.3 && 4.42 / 32.6 && 5.00 / 4.30 \\
\end{tabular}

\caption{Calculated values for the bulk modulus $K$ and
elastic constants of the BCC structure at several values of
$\lambda=\kappa a$, where $a$ is the average interparticle spacing,
for $n=0.23$ (first number) and $n=0.50$ (second number). All values
are in units of N/m$^{2}$.}
\label{elastictable}
\end{table}

While our framework assigns a surface charge that is smaller than the
value suggested in the original analysis~\cite{Sirota89}, it is in fact
quite common for the effective charge to be reduced when compared with the
net or titratable charge. To calculate the renormalized charge $Z^*$
and acid concentration $n^*_{HCl}$, Alexander {\sl et
al.}~\cite{Alexander84} employed a Wigner-Seitz approximation similar
in principle to our way of determining the bulk free energies of the
various lattices. Their method relies on the fact that in interstitial
regions, which occupy a large fraction of the system, the potential
$\Psi$ is relatively uniform and changes abruptly in a boundary layer
near the colloidal surfaces. As a result, the counterions are more
tightly constrained to locations near the surfaces of these colloidal
particles, and one can re-interpret these ions as part of an effective
sphere with a reduced charge. Again, since neither the effective
charge nor the local ion concentration can be measured directly in
experiments, the precise confirmation of theory has proven difficult.
Because our model allows the simultaneous prediction of the phase
boundary and the shear moduli, we can avoid these uncertainties.

There are, of course, additional interactions that we have neglected.  The
van der Waals attraction is much weaker than the screened Coulomb potential at
the interparticle spacing of $a\approx140$~nm.  We expect that
correlation effects such as overcharging should have minimal
consequence in a system of monovalent salt ions
of NaCl~\cite{Nguyen00}.  Dispersion forces are important at ion concentrations greater than
100 $\mu$mol~\cite{Ninham}.  However, this effect occurs at distance scales
on the order of 10 $\mu$m, and we do not expect this to change
our model. 

Finally, we note that in the calculation of the bulk free energy, we
employ free-volume theory even in the relatively low-particle-density
regime -- a regime in which pure hard-core interactions would predict
a fluid phase. However, the screened Coulomb potential stabilizes the
lattice structures and, in turn, lowers the melting density of the
system. One could characterize the system with an
effective density of higher magnitude, determined by the
Barker-Henderson effective diameter of particles~\cite{Gast83}.  A scheme like this would
introduce another parameter into our model which would divide the
screened-Coulomb potential into a ``hard'' part and a ``soft'' part.

We have established a geometrical framework for
understanding the structural and mechanical properties of charged
colloids. We have connected the phase
diagram to the elastic moduli of the system with only one adjustable
parameter, the surface charge. The balance between the drive of the
system's entropy for maximal packing fraction and the need of its
screened repulsion for minimal interfacial area accounts for most of
the properties found in experiments. Its elucidation can serve as
an intuitive guide to the engineering of all such charged systems.

We gratefully acknowledge stimulating conversations with
T.C.~Lubensky, T.R.~Powers, M.O.~Robbins, D.A.~Weitz, and A.G.~Yodh. This work was supported by NSF Grant DMR97-32963, the Donors of the Petroleum Research
Fund, administered by the American Chemical Society, the University of
Pennsylvania Research Foundation, and a gift from L.J.~Bernstein.
R.D.K. was also supported by the Alfred P. Sloan foundation.


\begin{references}
\bibitem{Hachisu73} S. Hachisu, Y. Kobayashi, and A. Rose, J. Col.
Int. Sci. {\bf 42}, 342 (1973).
\bibitem{Yoshiyama86} T. Yoshiyama, Polymer {\bf 27}, 828 (1986).
\bibitem{Russel89} W. B. Russel, D. A. Saville, and W. R. Schowalter,
{\sl Colloidal Dispersions} (Cambridge University Press, New York, 1989).
\bibitem{Mono89} Y. Monovoukas and A. P. Gast, J. Col. Int. Sci. {\bf
128}, 533 (1989).
\bibitem{Robbins88} M. O. Robbins, K. Kremer, and G. S. Grest, \jcp
{\bf 88},3286 (1988).
\bibitem{Ziherlshort} P. Ziherl and R. D. Kamien, \prl {\bf 85}, 3528
(2000).
\bibitem{Ziherllong} P. Ziherl and R. D. Kamien, J. Phys. Chem. B {\bf
105}, 10147 (2001).
\bibitem{Sirota89} E. B. Sirota, {\sl et al.},
\prl {\bf 62}, 1524 (1989).
\bibitem{Grier98} J. A. Weiss, A. E. Larsen, and D. G. Grier, \jcp
{\bf 109}, 8659 (1998).
\bibitem{Brenner} T. M. Squires and M. P. Brenner, \prl {\bf 85}, 4976 (2000).
\bibitem{Balagurusamy} V. S. K. Balagurusamy, 
G. Ungar, V. Percec, and
G. Johansson, 
{\it J. Am. Chem. Soc.} {\bf 119}, 1539 (1997).
\bibitem{Thomson1887} W. Thomson, Phil. Mag. {\bf 24}, 503 (1887).
\bibitem{Robbinspers} We thank M. O. Robbins for discussions on this point.

\bibitem{Gast83} A. P. Gast, C. K. Hall, and W. B. Russel, Faraday
Discuss. Chem. Soc. {\bf 6}, 189 (1983).
\bibitem{Joanny} J. F. Joanny, J. Col. Int. Sci. {\bf 71}, 622
(1979).
\bibitem{Lubenskytext} P. M. Chaikin and T. C. Lubensky, {\sl
Principles of Condensed Matter Physics} (Cambridge University Press,
New York, 1980).
\bibitem{Brakke} K. Brakke, Exp. Math. {\bf 1}, 141 (1992).
\bibitem{Alexander84} S. Alexander, 
P. M. Chaikin, P. Grant,
G. J. Morales, P. Pincus, and D. Hone, 
\jcp {\bf 80}, 5776 (1984).

\bibitem{Nguyen00} T. T. Nguyen, A. Yu. Grosberg, and
B. I. Shklovskii, \prl {\bf 85}, 1568 (2000).
\bibitem{Ninham} M. Bostr\"om, D. R. M. Williams, and B. W. Ninham, \prl {\bf 87}, 168103 (2001).

\end{references}
\end{document}